\documentclass[a4paper,fleqn,usenatbib]{mnras}
\usepackage[T1]{fontenc}
\usepackage{ae,aecompl}
\usepackage{amssymb}
\usepackage{graphicx}                            % Graphics package
\usepackage{amsmath} 
\usepackage{perpage} %the perpage package
\usepackage{color}
\usepackage{cleveref}
\usepackage{epsfig}
\usepackage{epstopdf}
\usepackage{lipsum}
\usepackage{array}
\newcolumntype{L}{>{\centering\arraybackslash}m{1.5cm}}
\usepackage{ctable}
\usepackage{tablefootnote}
\usepackage{footnote}
\newcommand*\colvec[3][]{\begin{pmatrix}\ifx\relax#1\relax\else#1\\\fi#2\\#3\end{pmatrix}}

\newcommand\kls[1]{{{#1}}}
\newcommand{\beq}{\begin{equation}}
\newcommand{\eeq}{\end{equation}}
\newcommand{\beqn}{\beq \begin{aligned}}
\newcommand{\eeqn}{\end{aligned}\eeq}

\newcommand{\msun}{\, M_\odot}

\newcommand{\mbh}{M_{\rm bh}}

\newcommand{\mbulge}{M_{\rm bulge}}

\newcommand {\gtsim} {\ {\raise-.5ex\hbox{$\buildrel>\over\sim$}}\ }
\newcommand {\ltsim} {\ {\raise-.5ex\hbox{$\buildrel<\over\sim$}}\ } 
\newcommand{\mchirp}{\mathcal{M}}

\title[Individual SMBH Binary Constraints from PTA Limits]{Constraints on Individual Supermassive Black Hole Binaries from
  Pulsar Timing Array Limits on Continuous Gravitational Waves}
\author[Schutz \& Ma]{
Katelin Schutz,$^{1}$\thanks{kschutz@berkeley.edu}
Chung-Pei Ma,$^{2}$
\\
$^{1}$Department of Physics, University of California at Berkeley, Berkeley CA 94720\\
$^{2}$Department of Astronomy, University of California at Berkeley, Berkeley CA 94720}

\date{\today}
\pubyear{2015}

\begin{document}
\label{firstpage}
\pagerange{\pageref{firstpage}--\pageref{lastpage}}
\maketitle

\begin{abstract}
  Pulsar timing arrays (PTAs) are placing increasingly stringent
  constraints on the strain amplitude of continuous gravitational waves
  emitted by supermassive black hole binaries on subparsec scales.  In this
  paper, we incorporate independent %measurements of 
  {information about} the dynamical masses
  $\mbh$ of supermassive black holes in specific galaxies at known
  distances and %leverage 
  {use} this additional information to further constrain
  whether or not those galaxies could host a detectable supermassive black
  hole binary. We estimate the strain amplitudes from individual binaries
  as a function of binary mass ratio for two samples of nearby galaxies:
%that are most likely to host an individually-resolvable
%  supermassive black hole binary
  (1) those with direct dynamical measurements of $\mbh$ in the literature,
  and (2) the 116 most massive early-type galaxies (and thus likely hosts
  of the most massive black holes) within 108 Mpc from the MASSIVE Survey.
  Our exploratory analysis shows that the current PTA upper limits on
  continuous waves \kls{(as a function of angular position in the sky)} can already constrain the mass ratios of
  hypothetical black hole binaries in {many} galaxies in our samples.  The
  constraints are stronger for galaxies with larger $\mbh$ and at smaller
  distances.  For the black holes with $\mbh \ga 5\times 10^9 M_\odot$ at
  the centers of \kls{NGC 1600}, NGC 4889, NGC 4486 (M87) and NGC 4649 (M60), any binary
  companion in orbit within the PTA frequency bands would have to have a
  mass ratio of {a few percent or less}.%about 1:10.
% We argue that the galaxies considered in
%  this paper are , and we find that pulsar timing already places
%  constraints on the existence of a binary in several galaxies.
  \end{abstract}


%%%%%%%%%%%%%%%%%%%%%%%%%%%%%%%%%%%%%%%%%%%%%%%%%%

%%%%%%%%%%%%%%%%% BODY OF PAPER %%%%%%%%%%%%%%%%%%

%%%%%%%%%%%%%%%%%%%%%%%%%%%%%%%%%%%%%%%%%%%%%%%%%%%%%%%%%%%%%%%%%%%%%%%%%%%%%%%%%%%%%%%%%%%%%%%%%%%%%%%%%%%%%%%%%%%%%%%%%%%%%%%%%%
\section{Introduction}

Pulsar timing arrays (PTAs) are monitored collections of millisecond
pulsars, which are rapidly spinning magnetized neutron stars that emit pulses with an extremely precise tempo.
%Effectively, these pulsars can function as astrophysical clocks.%which are magnetized, rapidly rotating neutron stars. Millisecond
%pulsars emit pulses at regular intervals on the order of $10^{-3}$ s, and
%the stability of these pulse emissions rivals the stability of atomic
%clocks owing to the pulsars' large angular momentum. 
Pulsar timing is sensitive to gravitational waves (GWs), which affect the relative arrival time of pulses as
measured by observers. Thus, these pulsars act as exquisite probes of GWs
and their underlying sources %\citet{1983ApJ...265L..39H},
\citep{1978SvA....22...36S, 1979ApJ...234.1100D, 1983ApJ...265L..39H,
  1989ASIC..262..113R,1990ApJ...361..300F}. %\citealt{1983ApJ...265L..39H}
Three PTAs are currently in operation: the North American Nanohertz
Observatory for Gravitational waves (NANOGrav;
\citealt{2009arXiv0909.1058J}), the Parkes Pulsar Timing Array (PPTA;
\citealt{2013PASA...30...17M}), and the European Pulsar Timing Array (EPTA;
\citealt{2008AIPC..983..633J}).  PTAs are sensitive to GW frequencies
% is determined by the Nyquist frequency of observations and by the total
% length of observations: this frequency range
roughly between 1-100 nHz, a range that depends on the total observational
time and observing cadence.

The strongest anticipated source of gravitational waves in the PTA
frequency band is an as-of-yet undetected population of supermassive black
hole (SMBH) binaries
\citep{1980Natur.287..307B,2003ApJ...595..614W,2003ApJ...583..616J}. 
Under the standard hierarchical picture of galaxy assembly,
these SMBH binaries are predicted to form near the centers of galaxies via
dynamical friction after galaxy-galaxy mergers, and are expected
to be relatively common in a cosmological sense \citep{2004ApJ...611..623S}. In spite of the
theoretical ubiquity of SMBH binaries, many of their properties are still
uncertain, such as their formation history \citep{2003ApJ...582..559V} and
their dynamical influences on host galaxies \citep{2009ApJS..182..216K}.

One expects the whole population of SMBH binaries to emit GWs that add
incoherently, forming a roughly isotropic stochastic background of
gravitational waves
\citep{1995ApJ...446..543R,2001astro.ph..8028P,2008MNRAS.390..192S}.  Upper
limits on the stochastic background are becoming tighter as the timing data
and analysis algorithms improve \citep{%2013Sci...342..334S,
  2015arXiv150803024A, 2015MNRAS.453.2576L,Shannonetal2015}. In addition to
the stochastic background, there may be individual sources that emit
sufficiently strong continuous GWs to overpower the background. Such
individual sources would require relatively nearby massive black hole
binaries to exceed the signal of the GW background
\citep{2009MNRAS.394.2255S, 2010CQGra..27h4016S}.
% we stress that these are the most important criteria for eventually
% detecting continuous GWs using PTAs. To date, no individual sources of
% continuous gravitational waves in the PTA band have been detected
% \kls{cite all the collaborations}. In fact, as previously alluded to,
% there has been no confirmed detection of any SMBH binary as of yet, though
% there are some potentially promising candidates
% \kls{cite{possiblebinaries}}.

%Though much of the effort of detecting gravitational waves has been focused
%toward detecting the GW background, recent study has shown that the first
%gravitational waves detected by PTAs could conceivably come from individual
%sources. As \kls{cite sesana paper} has shown, the probability of the first
%detection of GWs being continuous (as opposed to the first detected GWs
%coming from the stochastic background) is around $\sim$10\%, which is much
%higher than previously thought. This result may encourage some shift in PTA
%observing strategies, which we hope to address in this paper. Additionally,
%the idea of detecting an individual source of continuous gravitational
%waves is quite enticing because it may provide the first opportunity to do
%in-depth multi-messenger astronomy with a particular source. \kls{insert
%  extra motivation if I can think of some CW vs GWB motivation}

Thus far, each of the three PTA teams has published an analysis of the upper
limit on GWs from individual binary sources.  From a sample of 17 pulsars,
the NANOGrav 5-year data \citep{Demorestetal2013} placed an \kls{all-sky} upper limit on
the strain amplitude of $h_0 < 3.0\times 10^{-14}$ (95\%
confidence level) at 10 nHz \citep{2014ApJ...794..141A}.  PPTA reported an
upper limit of $h_0 < 1.7 \times 10^{-14}$ at 10 nHz
\citep{2014MNRAS.444.3709Z} based on observations of 20 pulsars from their
data release DR1 in \citet{Manchesteretal2013}.  EPTA used data from \kls{42}
pulsars and their Bayesian pipeline (assuming the sources do not evolve on
the timescale of observations) gives an upper limit of $h_0 < 1.1 \times
10^{-14}$ at 10 nHz \citep{2015arXiv150902165B}.  These values were
marginalized over all %sky locations 
\kls{lines of sight} and can vary by a factor of a few at
different parts of the sky. \kls{These constraints, including their full dependence on angular position in the sky, can be used to probe binary black hole astrophysics.}

In this paper we present an exploratory study of the implications of these
constraints \kls{(as a function of angular position in the sky)} for specific known nearby galaxies that could host a detectable
SMBH binary.  Information about the total mass and location of possible SMBH binaries
provides additional constraining power to PTAs and sharpens the limits on
the binary properties.  For galaxies with such information, we estimate the
necessary PTA sensitivity for constraining the existence of a black hole binary.
%and obtain the constraints on whether some of these galaxies can
%host a binary.
  In particular, we examine two samples of galaxies: (1)
galaxies in which the masses of the central black holes have been
dynamically measured from luminous tracers such as stars, gas, and masers;
(2) galaxies in the MASSIVE Survey \citep{2014ApJ...795..158M}, which are
the most massive galaxies within $\sim$100 Mpc and whose SMBH masses can be
inferred by standard scaling relations with host galaxy properties.  Using
the measured black hole mass and distance to these sources, we estimate the
constraints on the binary mass ratio allowed by current PTA upper limits \kls{at the angular positions of our sample galaxies}.

The rest of the paper is organized as follows. In Section \ref{sec:model},
we discuss the physical parameters that can be constrained assuming a
signal model consisting of a SMBH binary with circular orbits evolving purely
under energy lost via the emission of gravitational waves. %The key
%parameters are the maximum GW amplitude for a given SMBH with a
%dynamically-known mass and the maximum mass ratio if the SMBH is actually a
%binary. 
In Section \ref{sec:samples} we describe the aforementioned samples of
galaxies whose SMBHs will be the focus for the rest of the paper.
% One sample consists of galaxies whose SMBHs have been dynamically
% measured. The other sample contains galaxies in the MASSIVE survey, which
% are the most massive galaxies within $\sim$100 Mpc and whose SMBH masses
% can be inferred by standard scaling relations with host galaxy
% properties.  Among these galaxies are the ones in the MASSIVE survey; we
% motivate these as being among the best candidates for hosting a
% detectable continuous GW source. In Section \ref{sec:h0} We infer the
% necessary PTA sensitivity for constraining the existence of SMBH binaries in our
% sample galaxies and find that PTAs can already place weak constraints on
% whether some of these galaxies can host a binary.
In Section \ref{sec:cons}, we explore constraints from PTAs on galaxies in
our samples and estimate limits on the mass ratios of {many} hypothetical
SMBH binaries.
%We find that the
%constraints on some SMBH binary mass ratios are so tight that the detectability
%of a binary is practically ruled out.
 %Given all this information, in Section
%\ref{sec:pulsar} we find sky locations where the addition of a new pulsar
%to the PTA {maybe NANOgrav rather than a generic PTA, since we know all
  %their pulsar info} would lead to further constriants on hypothetical
%SMBH binaries hosted by our sample galaxies. For the ``hottest'' pulsar hotspots,
%we show what timing precision is required {add more to this when there
  %is more development}. 
We discuss the results of this paper and future work in Section \ref{sec:con}.
%In Section {ref} we use the masses of dynamically measured SMBHs to infer the necessary sensitivity for constraining the possibility that the measured SMBH binary is actually a binary.
%%%%%%%%%%%%%%%%%%%%%%%%%%%%%%%%%%%%%%%%%%%%%%%%%%%%%%%%%%%%%%%%%%%%%%%%%%%%%%%%%%%%%%%%%%%%%%%%%%%%%%%%%%%%%%%%%%%%%%%%%%%%%%%%%%%%%%%%%%%%%%%%%%%%%%%%%%%%%%%%%%%%%%%%%%%%%%%%%%%%%%%%%%%%%%%%%%%%%%%%%%%%%%%%%%%%%%%%%%%%%%%%%%%%%%%%%%%%%%%%

%\section{Galaxies of Interest}
%\section{Constraints on Mass Ratios of Potential Black Hole Binaries}
%\section{Chirp Mass $\cal M$ vs Dynamical Mass $\mbh$}
\section{Constrained Physical Parameters}
\label{sec:model}

%We begin by motivating the consideration of particular galaxies for the
%rest of the paper. Intuitively, in order for a continuous GW to be
%detectable, it must emit very strong gravitational waves and be relatively
%close to us as observers on earth. More quantitatively, 

The amplitude of continuous gravitational waves can be parameterized by a
dimensionless strain amplitude, $h_0$. The GW signal depends on other
physical characteristics of a SMBH binary such as the binary inclination
and phase, but PTA constraints are reported in terms of this intrinsic
amplitude \kls{with the other binary characteristics marginalized over}. %Following \citet{2012PhRvD..85d4034B},
For a SMBH binary at \kls{leading} post-Newtonian order under the
assumption of circular orbits and evolution purely by energy loss via
gravitational radiation, the strain amplitude is given by
%
%\begin{eqnarray}
\beq
h_0= 2.76\times 10^{-14}  \left( \frac{{\cal M}}{10^9 M_\odot} \right)^{5/3}
\left(\frac{10\,\rm Mpc}{d_L}\right)\left(\frac{f}{10^{-8}{\rm Hz}}\right)^{2/3} 
  \,,
%\nonumber
%&& h_0 = 2\, \frac{(G \mathcal{M} )^{5/3} (\pi f_{gw})^{2/3}}{c^4 \,d_L} \\
%&=& 2.76\times 10^{-14} \left( \frac{{\cal M}}{10^9 M_\odot} \right)^{5/3}
% \left(\frac{10\,\rm Mpc}{d_L}\right)\left(\frac{f}{10^{-8}{\rm Hz}}\right)^{2/3} 
% \,, \nonumber
\label{h0}
\eeq
%\end{eqnarray}
%
where $f$ is the frequency of the emitted gravitational waves, $d_L$ is
the luminosity distance to the source, and $\mchirp$ is the chirp mass of a
binary system consisting of two black holes of mass $m_1$ and $m_2$, with a
total mass $\mbh =m_1 + m_2$ and mass ratio $\kls{q} \equiv m_2/m_1 \le 1$:
\beq \mchirp \equiv
\frac{(m_1 m_2)^{3/5}}{(m_1+m_2)^{1/5}} = \mbh \frac{\kls{q}^{3/5}}{(1+\kls{q})^{6/5}} \,.
\label{massratio}
\eeq 
Since $\kls{q} \le 1$ by definition, it follows that 
\beq
 {\cal M}  \le 2^{-6/5} \mbh \approx 0.435 \mbh \,.
\label{maxChirp}
\eeq
% but we can incorporate independent knowledge about distances and masses,
% with the mass being the dominant factor in the strain amplitude.  If we
% assume a fiducial GW frequency of $10^{-8}$ Hz, then we can know the
% strain amplitude of a source provided that we know the distance to that
% source and its chirp mass.  Since there have been no confirmed detections
% of a SMBH binary, there is no way for us to know the chirp mass \emph{a
%   priori}.  However,
Thus, for a given total black hole mass $\mbh$, the chirp mass is at a maximum for an
equal-mass binary and decreases monotonically as the binary mass ratio
$m_2/m_1$ decreases. The chirp mass is zero for a single black hole.
% when $m_1 = m_2$ with a value of $2^{-6/5} (m_1+m_2)$. 
%Assuming this chirp mass will set a maximum value of $h_0$
%where we can begin to constrain the existence of a SMBH binary in a particular
%galaxy.

Combining Eqs.~(\ref{h0}) and (\ref{maxChirp}), we obtain
the equal-mass strain amplitude
\beq
 %h_0 \le 
 h_{0,\rm eq} = 
 6.9\times 10^{-15}\left( \frac{\mbh}{10^9 M_\odot} \right)^{5/3}
 \left(\frac{10\,\rm Mpc}{d_L}\right)\left(\frac{f}{10^{-8}{\rm Hz}}\right)^{2/3} \,.
\label{maxh0}
%\frac{\cal{M}}{10^9 M_\odot} = \left(\frac{h_0}{2.76 \times 10^{-14}}\right)^{3/5} \left(%\frac{d_L}{10~\text{Mpc}}\right)^{3/5} \left(\frac{f_{\rm gw}}{10^{-8}~\text{Hz}}\right)^{-2/5}  \,.
%\label{chirpmass}
\eeq 
Since the GW strain amplitude $h_0$ scales monotonically with the chirp
mass, $h_{0,\rm eq}$ corresponds to the maximum allowed value of $h_0$ for
GWs at frequency $f$ emitted by a SMBH binary with total mass $\mbh$ at a
distance of $d_L$.  This theoretical maximum on the value of $h_0$
represents the most optimistic limit that PTAs must achieve before being
able to detect continuous GWs from SMBH binaries in specific galaxies.
%Since we can already approximately rule out the maximum
%theoretically-allowed values of $h_0$ for several objects, we are thus able
%to constrain the parameter space of mass ratios for a hypothetical
%binary. With the fiducial, sky-averaged constraint on $h_0$, one can arrive
%at the corresponding constraint on the chirp mass:
%For a given upper limit of $h_0$ at the location of a potential GW source
%at distance $d_L$, we arrive at a corresponding upper limit on the chirp mass

%
In the next two sections, we will consider galaxies whose central black
holes have a total mass $\mbh$ that either have been directly measured or
can be inferred statistically from their host galaxy properties by using
empirical scaling relations.
%For such systems with known $m$,
%an upper limit on (and hence $\cal M$) can provide an upper limit 
%on the mass ratio of the candidate binary, 
%As we discuss in Sec~3, 
These dynamical mass measurements 
%of supermassive black holes at centers of galaxies
constrain the total mass $\mbh$ but {in general} cannot distinguish a single black hole from a binary.
%  Our main goal in
%this paper is to combine the latest dynamical black hole measurements of
%$m$ and PTA upper limits on $h_0$ and use Equations~(\ref{chirmass}) and
%(\ref{massratio}) to place constraints on the mass ratio of candidate
%binary black holes allowed in these galaxies.
%As we show in Section \ref{sec:cons}, 
Several black holes in our sample have sufficiently large $\mbh$ and small $d_L$
that the values of $h_{0,\rm eq}$ from Eq.~(\ref{maxh0}) exceed the current
PTA upper limits on $h_0$.  For these objects, we can further constrain the
allowed mass ratio $m_2/m_1$ of potential binary black holes using the PTA
detection limits within the detectable orbital frequencies.
To obtain such constraints, we note that
for a given $d_L$, $f$, and PTA upper limit on $h_0$, Eq.~(\ref{h0})
provides an upper limit on the chirp mass, which we refer to as ${\cal M}_{\rm PTA}$.
We can then use Eq.~(\ref{massratio}) to estimate the upper limit on $m_2/m_1$:
\beq
\frac{m_2}{m_1} \le \frac{1 - \sqrt{1 - 4 ({\cal M}_{\rm PTA}/\mbh)^{5/3}}} {1 +
  \sqrt{1 - 4 ({\cal M}_{\rm PTA}/\mbh)^{5/3}}} \,.
%\frac{m_2}{m_1} = \frac{m - \sqrt{m^2 - 4\, \mathcal{M}^{5/3} m^{1/3}}}{m + \sqr
%t{m^2 - 4\, \mathcal{M}^{5/3} m^{1/3}}}, 
\label{ratio}
\eeq
We examine the black hole data and present the results in the next two sections.

\begin{figure*}
\includegraphics[width=0.48\textwidth]{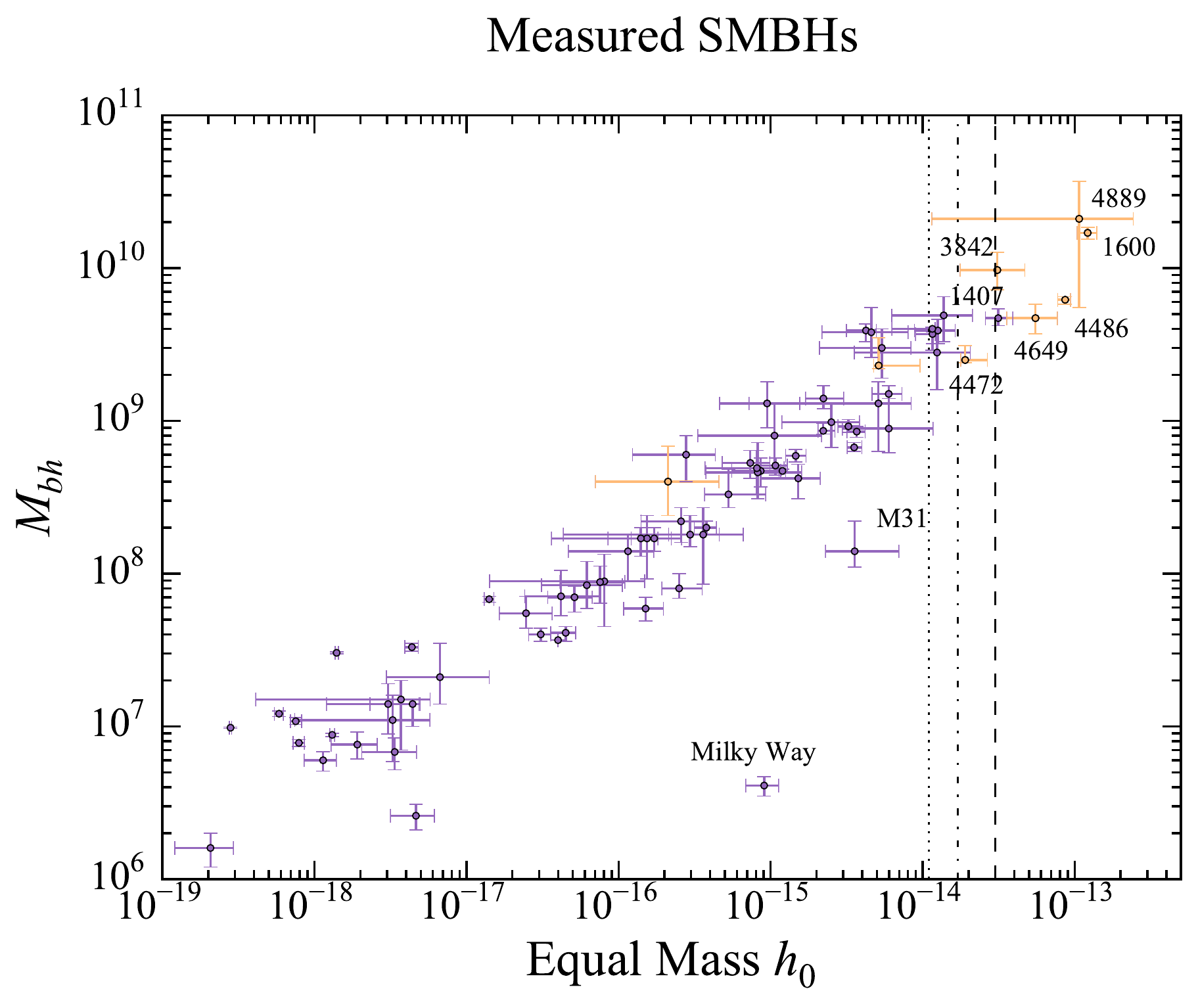}
%\vspace{0.2in}
\includegraphics[width=0.48\textwidth]{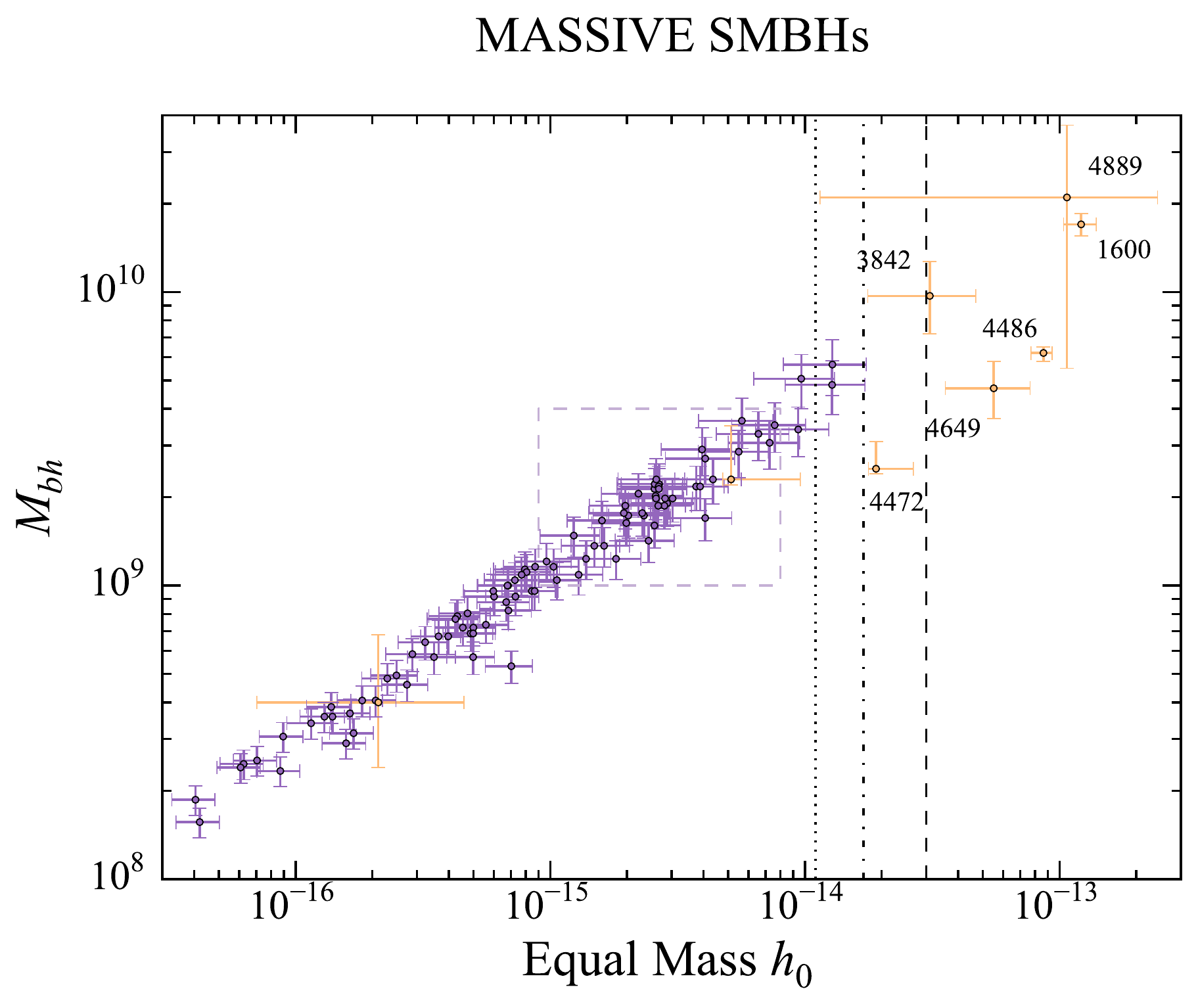}
\caption{(Left) Black holes with dynamical mass measurements $\mbh$ and the
  strain amplitudes $h_{0,\rm eq}$ of GWs that would be emitted if the
  sources were in equal-mass binaries.  The vertical lines represent the
  \kls{all-sky} upper limits on $h_0$ from continuous waves by EPTA
  (dotted), PPTA (dot-dashed), and NANOGrav (dashed). Here $h_0$ is
  determined at a {GW} frequency of 10 nHz.  (Right) Same as the left panel but
  for the sample of 116 most massive early-type galaxies within 108 Mpc
  targeted by the MASSIVE Survey.  The purple points here show black hole
  masses estimated from the $\mbh$-$\sigma$ relation, where $\sigma$ is the
  measured stellar velocity dispersion of the galaxy (see Section \ref{sec:massive}).  The
  dashed box shows the region spanned by the MASSIVE galaxies when the
  $\mbh$-$\mbulge$ relation is instead used to infer $\mbh$. The range of
  $\mbh$ based on $\mbulge$ is smaller because the MASSIVE galaxies are
  selected based on their stellar masses and not velocity dispersions.  The
  \kls{eight} orange points in both panels represent the galaxies common to both
  samples, i.e. the MASSIVE galaxies with dynamically measured $\mbh$.}
\label{fig:maxh0}
\end{figure*}

\begin{figure*}
%\center{\includegraphics[width=2.5in]{sigma.pdf}{bulge.pdf}}
 \includegraphics[width=\textwidth]{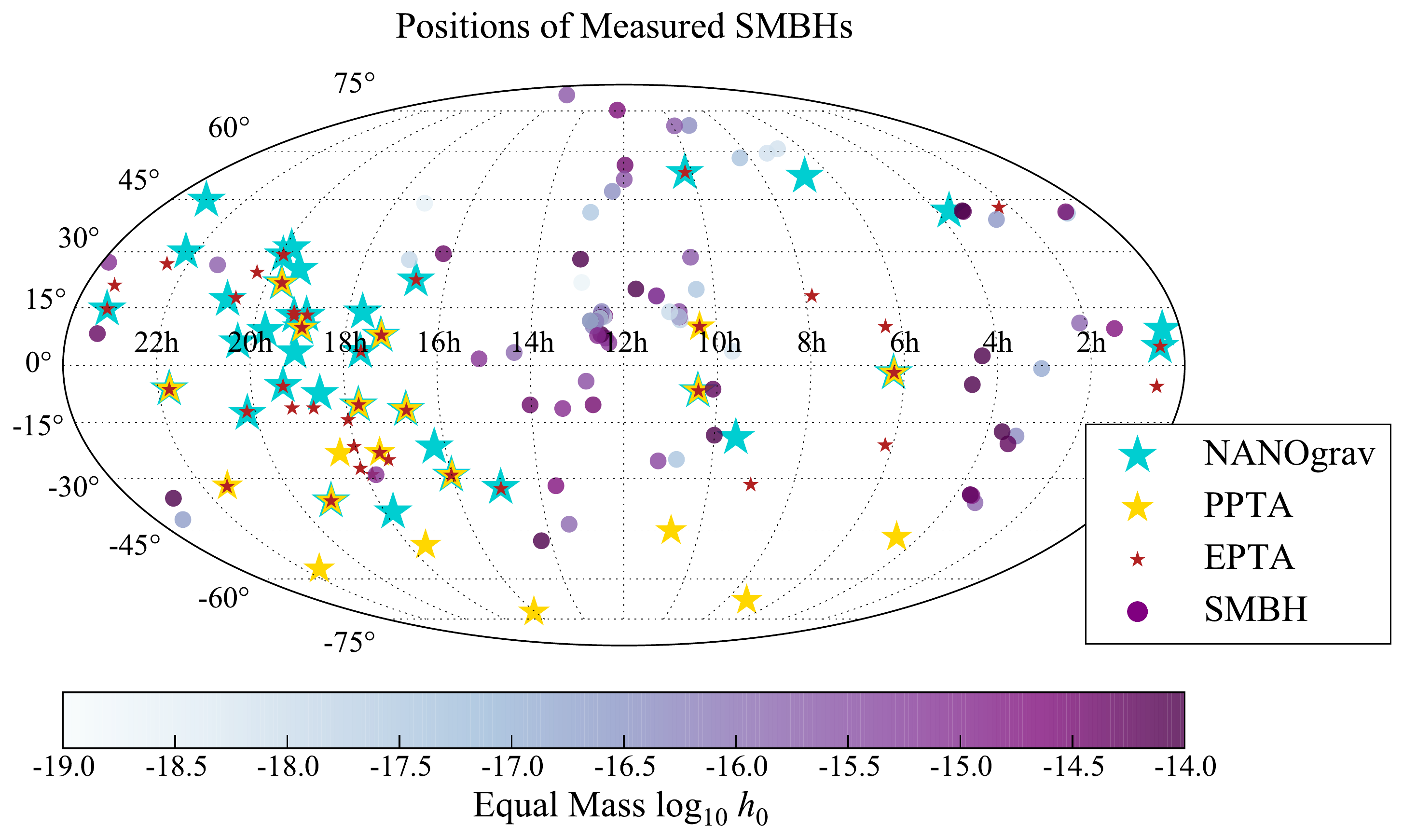}
%\includegraphics[width=3in]{bulge.pdf}
%\\\includegraphics[width=0.42\textwidth]{sigma.pdf}\\\includegraphics[width=0.42\textwidth]{bulge.pdf}}
\includegraphics[width=\textwidth]{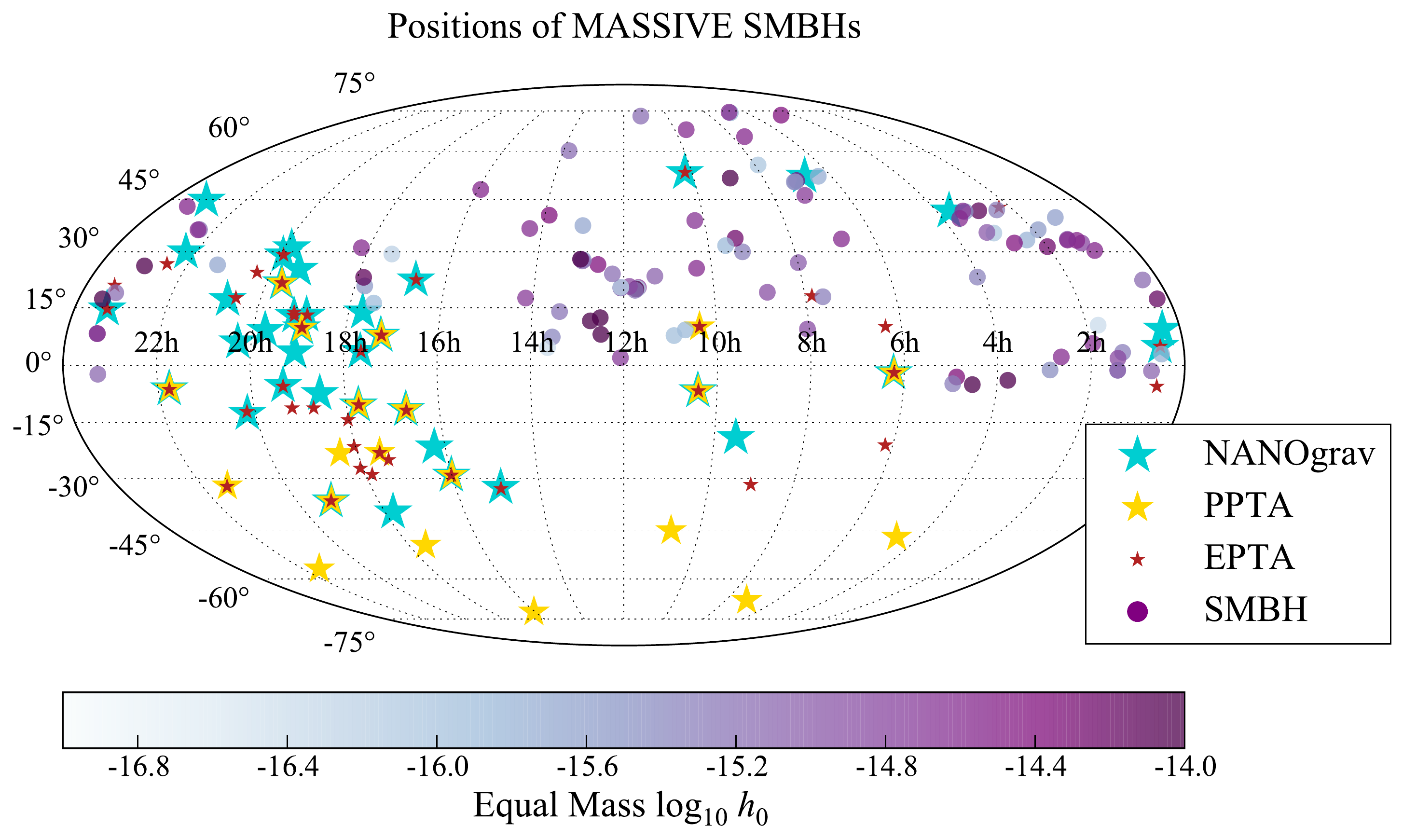}
%\includegraphics[width=3.5in]{bulgemap.pdf}
%\\\includegraphics[width=0.42\textwidth]{sigma.pdf}\\\includegraphics[width=0.42\textwidth]{bulge.pdf}}
\caption{%(Top) Same as Figure~1 but for the MASSIVE Survey.
%The dynamical mass $\mbh$ is based on scaling relations with two different properties of the host galaxy: stellar velocity dispersion and the $\mbh-\sigma$
%relation (left); galaxy bulge mass and the $\mbh-\mbulge$ relation (right).
%Orange points in the top
%  panel represent SMBHs with measured masses which happen to also be in the
%  MASSIVE survey. The red and blue boxes in the top panel represent the
%  domains of the middle and bottom panels, to illustrate where the MASSIVE
%  points would lie relative to the measured ones.  
%The vertical lines
  %represent upper bounds on $h_0$ from
  %NANOgrav (solid), PPTA (dashed), and the dotted line shows a
  %factor of 2 improvement on PPTA, which is anticipated for the latest N-year data.
% which is anticipated for the NANOgrav 9-year data.
  (Top) \kls{Angular positions} of the supermassive black holes with dynamical mass
  measurements (purple circles) and the pulsars with timing measurements
  from NANOGrav (cyan stars), PPTA (yellow stars), and EPTA (red stars).
  The shading of the purple circles represents the {maximum} strain amplitudes
  $h_{0,\rm eq}$ of GWs (at 10 nHz) given by Eq.~(4) that would be emitted
  if the sources were in equal-mass binaries.  (Bottom) Same as the top
  panel but for the sample of 116 most massive early-type galaxies within
  108 Mpc targeted by the MASSIVE Survey.  The \kls{angular positions} of the
  strongest \kls{potential GW} sources will influence the comparison between PTAs
  with different pulsars: due to the steep angular
  dependence of the PTA antenna function \citep{2010arXiv1008.1782C}, the
  most promising sources to be constrained by future data analysis are dark
  purple circles that lie along similar lines of sight as the {best-timed} PTA pulsars.}
\label{fig:map}
\end{figure*}

%\begin{figure*}[!]%[htb]
%\center{\includegraphics[width=0.48\textwidth]{measuredmap.pdf}\\\includegraphics[width=0.48\textwidth]{sigmamap.pdf}}
%\\\includegraphics[width=0.48\textwidth]{bulgemap.pdf}}
%\center{\includegraphics[width=6.in]
%{measuredmap.pdf}}
%\caption{Sky map of the locations of 74 black holes with dynamically
%  measure masses (purple circles) and the pulsars with timing measurements
%  from NANOgrav (cyan stars), PPA (red stars), and EPTA (yellow
%  stars). Also shown are the sources' maximum strain amplitudes. The
%  spatial distribution of the strongest possible sources will influence the
%  comparison between PTAs with different spatial distributions of pulsars
%  and different spatial distributions of pulsars' timing residuals.}
%\label{fig:maps}
%\end{figure*}

\section{Two samples of black holes}
\label{sec:samples}
\subsection{Black holes with dynamical mass measurements}

We first examine a collection of \kls{77} galaxies whose SMBH masses have been
measured directly from kinematics of dynamical tracers within the
gravitational sphere of influence of the black holes.  These measurements
represent over two decades of observational efforts by various
collaborations and are obtained from high-angular-resolution spectroscopy
of the centers of galaxies within a distance of $\sim 120$ Mpc from earth.
Stars and ionized gas are the most commonly used tracers. Masers and molecular
CO gas with organized rotations have been used for a small set of black
holes. 
%\{CP said to elaborate... What was meant by this?}
We take the sample of 72 compiled in \citet{2013ApJ...764..184M}, and add
\kls{five} measurements published since then: NGC 4526 \citep{2013Natur.494..328D,2013arXiv1303.0834G}, M60-UCD1
\citep{2014Natur.513..398S}, NGC
1277 \kls{\citep{2015arXiv151104455W}}, NGC 1271 \citep{2015ApJ...808..183W}\kls{, and NGC 1600 \citep{nature1600}}.
%\citep{2015MNRAS.452.1792Y}.

These dynamical measurements are sensitive to the total enclosed mass
within the angular resolution scale allowed by the instruments and
observing conditions at the time the data were taken. With the exception of the 
center of the Milky Way \citep{2010RvMP...82.3121G}, the data can not
distinguish a single black hole from a binary below this spatial scale.  If
any one of these galaxies indeed harbors a SMBH binary, their relatively small
distances from earth and large masses make them the most promising sites
for continuous GW detections by the PTAs. Conversely, an upper limit on GW
strain amplitudes from PTAs provides a corresponding upper limit on the
allowed binary mass ratio given by Eq.~(\ref{ratio}).

We depict these data by showing the equal-mass $h_0$ from Eq.~\eqref{maxh0}
as a function of the measured SMBH mass in the left panel of
Figure~\ref{fig:maxh0}.  This value of $h_0$ corresponds to the largest
possible strain for a source with known total $\mbh$ and distance.  We find
a tight relationship as expected, since the scaling with mass is the
strongest power law in determining the value of $h_0$.  The black holes in
the Milky Way and M31 are extreme outliers due to their small
distances. 
%Thus, one can roughly see what kinds of total SMBH binary masses will
%be constrained by different upper limits on $h_0$. 
We show the \kls{angular}
distribution of these possible GW sources relative to the pulsars in
various PTAs in the top panel of Figure \ref{fig:map}; the \kls{angular}
distribution of sources and pulsars is crucial for
detectability because of the PTA antenna pattern. \kls{For instance, potential sources which lie along similar lines of sight as the best-timed pulsars will be the most constrained by PTAs.}  %We leave full PTA data
%analysis of the spatial dependence of detectability to future work.
%If the SMBHs in these galaxies actually turn out to
%be binaries, the measured dynamics would be sensitive to the total mass,
%$m_1 + m_2$. Thus, we have some information about the chirp mass,
%$\mchirp$, in addition to knowing the distance of these galaxies.

\subsection{The MASSIVE Survey: Most massive black holes within 100 Mpc}
\label{sec:massive}

The galaxies with dynamical $\mbh$ measurements discussed in Sec~3.1
represent a heterogeneous collection of both late-type and early-type
galaxies over a wide mass range.  The data points shown in the left panel of
Figure~\ref{fig:maxh0} are therefore a compilation of measurements obtained
with differing methods and selection criteria and do not represent results
from any well-defined survey or systematic search.  

For a more complete census of possible sites for the most massive black
holes in the local volume, we consider here the galaxies in the ongoing
MASSIVE survey (Ma et al 2014).  The MASSIVE survey is a volume-limited,
multi-wavelength spectroscopic and photometric survey of the 116 most
massive early-type galaxies within 108 Mpc in the northern sky (above a
declination of $-6^\circ$).  The galaxies in the survey are selected based
on stellar mass, and the survey is complete to an absolute $K$-band
magnitude of $M_K = -25.3$ mag, corresponding to a stellar mass of
$M*\approx 10^{11.5}\msun$.  The MASSIVE galaxies reside in diverse
environments. Despite their large stellar masses, only 9 of the 116
galaxies reside in the three well-known clusters Virgo, Perseus and Coma.
A total of 26 MASSIVE galaxies are relatively isolated and considered
``groupless'' according to the group catalog of \citet{Crooketal2007},
containing fewer than three group members.
%  This ongoing survey aims to address a wide range of outstanding
%problems in formation of massive galaxies, including the variation in dark
%matter fraction and stellar initial mass function (IMF) within and among
%early-type galaxies, the connection between black hole accretion and galaxy
%growth, and the late-time assembly of galaxy outskirts.

\kls{Eight} galaxies in the MASSIVE sample have published black hole masses in
the literature: NGC 4486 \citep{2011ApJ...729..119G,
  2013ApJ...770...86W}, %(Gebhardt et al. 2011;Walsh et al. 2013),
NGC 4472 and NGC 7619 \citep{2013AJ....146...45R}%(Rusli et al. 2013b)
, NGC 4649 \citep{2010ApJ...711..484S}%(Shen \&Gebhardt 2010)
, NGC 3842 and NGC 4889 \citep{2011Natur.480..215M,
  2012ApJ...756..179M}%(McConnell et al. 2011a, 2012)
, NGC 7052
\citep{1998AJ....116.2220V}\kls{, and NGC 1600 \citep{nature1600}}.  %(van der Marel \& van den Bosch 1998)
These galaxies are located at the high end of the $\mbh$ and galaxy bulge mass
relation. The ongoing MASSIVE survey is expected to
provide at least 15 new measurements of $\mbh$.
% Early-type galaxies have grown in number and size by a factor of two
% since redshift $z\approx 1$,
%which is thought to happen via dissipationless merging and accretion
%{cite all the papers the MASSIVE survey cites}. Thus, not only are the
%galaxies in the MASSIVE survey the most massive and nearby, but they are
%also likely to have undergone a merger recently (in a cosmological sense)
%{again, ask CP because I'm kind of speculating a bit} and it is
%possible that they host a SMBH binary.

To estimate the dynamical value of $\mbh$ for the entire MASSIVE sample, we
follow the standard practice of using a galaxy's measured stellar velocity
dispersion $\sigma$ or bulge mass $\mbulge$ and converting them into $\mbh$
from well-established scaling relations between black hole and galaxy
properties.  We use $\sigma$ from %Table~3 of
\citet[Table 3]{2014ApJ...795..158M}, and we obtain the bulge mass from $M_K$ in the
same table and convert it to $\mbulge$ using $\log_{10} M_* = 10.58 - 0.44
(M_K + 23)$.  We assign the total stellar mass to the bulge since the
MASSIVE galaxies are all early-type galaxies and most of them are
elliptical galaxies.  

In the right panel of Figure~\ref{fig:maxh0} we show $h_{0,\rm eq}$ for
equal-mass binaries as a function of $\mbh$ inferred for MASSIVE galaxies,
using the $\mbh$-$\sigma$ relation of \citet{2013ApJ...764..184M}.  The range
of $\mbh$ and hence $h_0$ is smaller when using the $\mbh$-$\mbulge$ relation
(indicate by the dashed box) because the MASSIVE survey is selected based
on galaxy mass and not $\sigma$.  Even though $\mbulge$ and $\sigma$ of
galaxies are positively correlated, there is a large scatter in $\sigma$ at
a given stellar mass.  For the rest of the paper, we focus on masses
inferred by the $\mbh$-$\sigma$ relation.
%The upshot of this is that there are MASSIVE
%galaxies whose SMBHs are constrained by current PTA limits when using
%masses inferred by the $\mbh-\sigma$ relation, whereas the same cannot be
%said of using masses inferred by the $\mbh-\mbulge$ relation.

The \kls{angular positions} of the MASSIVE galaxies and PTA pulsars are plotted in
the bottom panel of Figure~\ref{fig:map}. 
% We note that there are more MASSIVE galaxies strong potential sources of
% GWs near the pulsars than in the top panel.  While a full determination
% of the spatial dependence of constraints will be left to future work,
We note that several MASSIVE galaxies are along almost the same line of
sight as several PTA pulsars, enhancing their detectability owing to the
steep angular dependence of the PTA antenna patterns
\citep{1987GReGr..19.1101W,2010arXiv1008.1782C}.
\\

\begin{figure}%[htb]
\includegraphics[width = 0.48\textwidth]{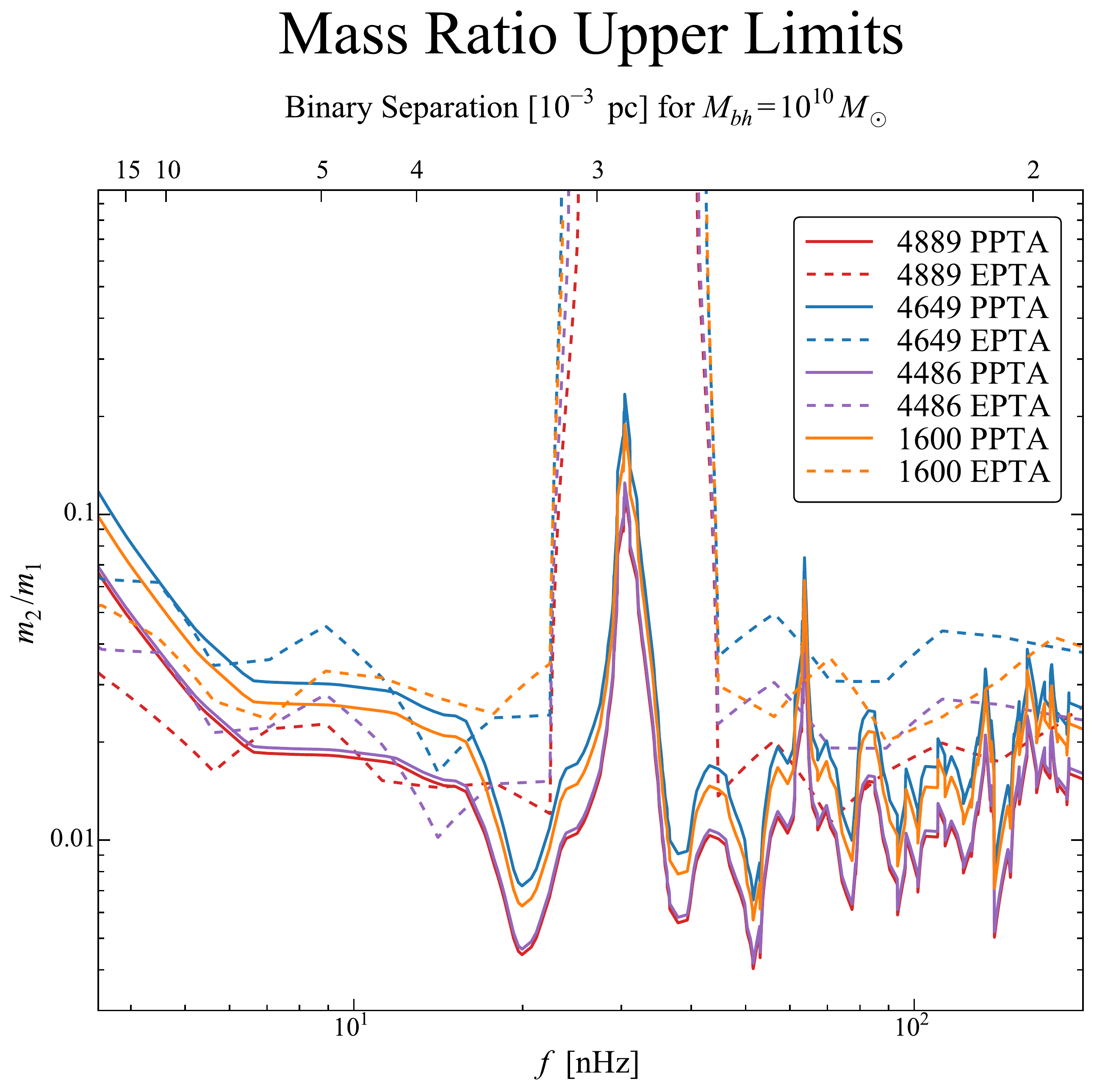}
\caption{{GW} frequency dependence of the upper limits on the black hole mass
  ratio ($m_2 \le m_1$ by definition) of hypothetical binaries located in
  the centers of NGC 4889 (red), NGC 4486 (purple), NGC 4649 (blue)\kls{, and NGC 1600 (orange)}.
  For each galaxy, the limit on $m_2/m_1$ is {computed} using the
  95\% upper limits on $h_0$ from EPTA (dashed curve) and PPTA
  (solid curve). Also shown is the
  corresponding binary orbital separation calculated assuming a $10^{10}
  \msun$ SMBH binary.  Here
  we use $\mbh$ for NGC 4486 from \citet{2011ApJ...729..119G} {and we assume that all SMBH masses are given by their mean measured values.}}
  %the
  %median value for NGC 4889 from \citet{2011Natur.480..215M}}.
% (the NANOGrav 5-year data are much less constraining and are therefore
% omitted.)  Shaded bands (some of which overlap) correspond to the
% 1-$\sigma$ uncertainties on the mass ratio coming from propagating the
% error on the dynamical mass measurements and from uncertainties on the
% local constraint on $h_0$.}
\label{fig:ratio_constraint}
\end{figure}

\section{Constraints on Individual Binary Mass Ratios}
\label{sec:cons}
%In our analysis below we will use these sky-averaged upper limits as fiducial
%constraints on strain amplitudes for the specific source locations we
%are interested in.  As we show, we can already begin to constrain the
%existence of SMBH binaries and their mass ratios in these galaxies. More spatially-accurate cons%traints will require a full MCMC analysis of actual PTA
%data (for instance using a pipeline such as the one discussed in \citealt{2013CQGra..30v4004E}), which will be the subject of future work.
%The GW frequency is usually not known \emph{a priori}.
%Following {cite nanograv etc}, we report our results at a fiducial GW
%frequency of $10^{-8}$ Hz.
%then we have all the information to know this
%maximum strain amplitude for a source hosted by any galaxy in our sample;
%we know the distance to that source and its total mass, which directly
%gives us the maximum possible value of $h_0$. 
%\begin{savenotes}
\begin{table*}
\label{tab:ratio} 
\begin{center}
\caption{\kls{Upper limits on the binary mass ratio, $m_2/m_1$, for SMBHs with measured total masses and for SMBHs in MASSIVE galaxies with their total masses inferred by standard scaling relations. We report limits from {EPTA and PPTA} constraint maps (taking into account the constraint at the angular position of each galaxy) at three representative {GW} frequencies. The constraints vary between the two PTAs depending on the angular position of the possible source and the frequency. Also listed are the total dynamical mass (measured or inferred), the distance, and the strain amplitude of GWs emitted in the scenario where each SMBH is in an equal-mass binary, which represents the maximum possible strain.}\label{tab}}
\begin{tabular}{l c c c c c c c c c } 
\toprule
%Galaxy & $\mbh$ & $d_L$  & $h_{\rm max}$ & \multicolumn{ 2}{ c }{PPTA} &\multicolumn{ 2}{ c }{NANOgrav} \\
Galaxy & $\mbh$ & $d_L$  & $h_{0,\rm eq}$ & \multicolumn{ 3}{ c }{$(m_2/m_1)_{\rm max}$ (EPTA)} & \multicolumn{ 3}{ c }{$(m_2/m_1)_{\rm max}$ (PPTA)} \\%&\multicolumn{ 3}{ c }{$(m_2/m_1)_{\rm max}$ (NANOGrav)} \\
\midrule
Measured         &  [$10^9 \,M_\odot$] & [Mpc] & [$10^{-14}$] &6 nHz& 10 nHz & 100 nHz &6 nHz& 10 nHz & 100 nHz \\%&6nHz& 10 nHz & 100 nHz \\
\midrule
%NGC 1600       & $1.7 \times 10^10$& 63.8 & $1.2\times 10^{-13}$ &  0.05 &0.02 &0.12& 0.04\\
NGC 1600 &  17& 63.8 & 12 & 0.026 & 0.032 & 0.022& 0.030 & 0.026 & 0.011 \\
NGC 4889     & $21$&102 & $11$& 0.020 & 0.019 & 0.018 & 0.021 & 0.018 & 0.008 \\%& 0.56 & 0.11 & -\\%& 0.04 & 0.02 & 0.11& 0.04\\ 
NGC 4486%(M87, stars)
& $6.2$  & 16.7 & $8.6$& 0.023 & 0.024 & 0.023 & 0.022 & 0.018 & 0.008 \\%& - & 0.14 & -\\%& 0.05 & 0.03 &0.14& 0.05\\
NGC 4486\footnotemark %(M87, gas) 
& $3.3$  & 16.7 & $3.0$ & 0.071 & 0.074& 0.071 & 0.069 & 0.058 & 0.026\\%&-&-&- \\
NGC 4649 %(M60)
& $4.7$  & 16.5 & $5.5$& 0.037 & 0.038 & 0.038 & 0.035 & 0.030 & 0.013  \\%& - & 0.28 & -\\%& 0.09 & 0.04 &0.28& 0.08\\
NGC 1407       & $4.7$  & 29.0 & $3.1$ & 0.11 & 0.12 & 0.072& 0.12 & 0.10 & 0.038\\%&-&-&-\\%& 0.19 & 0.08 &-&0.16\\
NGC 3842       & $9.7$  & 98.4 & $3.1$& 0.084 & 0.083 & 0.12 & 0.091 & 0.077& 0.030\\%&-&-&-\\%& 0.20 & 0.08 &-&0.17\\
NGC 4472 %(M49) 
& $2.4$  & 16.7 & $1.8$& 0.13 & 0.13 & 0.13& 0.13 & 0.10& 0.040 \\%&-&-&-\\% & 0.51 & 0.15 &-&0.37\\
NGC 1277  &4.9 & 71.0 & 1.4& 0.42 & 0.74 & - & -& - & 0.25 \\%&-&-&-\\
NGC 1550       & $3.8$  & 53.0 & $1.2$& - & - & 0.49& - & - & 0.15\\%&-&-&-\\% & -    %& 0.28 &-&-\\
IC 1459        & $2.8$  & 30.9 & $1.2$& 0.20 & 0.26 & 0.14& 0.21& 0.17 & 0.061\\%&-&-&-\\% & -    %& 0.29 &-&-\\
NGC 3091       & $3.7$  & 52.7 & $1.1$& 0.45 & 0.55 & -  & 0.30 & 0.23 & 0.077 \\%&-&-&-\\%& -    & 0.33 &-&-\\
NGC 5516       & $3.5$  & 60.1 & $0.92$  & 0.24 & 0.41 & 0.21& 0.15 & 0.13 & 0.047\\%&-&-&-\\%& -   &0.33  &-     &-\\
NGC 3115 & 0.89 & 9.5 & 0.60& - & - & - & - &- & 0.20\\
NGC 1332 & 1.5 & 22.7 & 0.60  & - & - & - & - &- & 0.31\\
NGC 1399 & 1.3 & 20.9 & 0.51& - & - & - & - &- & 0.30\\
NGC 7619 & 2.3 & 53.9 & 0.51 & - & - & - & - &- & 0.35\\
NGC 6086 & 3.8 & 139 & 0.46 &- & - & 0.68 & - &- & 0.19\\
A1836-BCG & 3.9 & 158 & 0.42 & - & - & - & - &- & 0.17\\
NGC 4594 & 0.67 & 10.0 & 0.35 &- & - & - & - &- & 0.28\\
NGC 4374 & 0.92 & 18.5 & 0.32& - & - & - & - &- & 0.51\\
A3565-BCG & 1.4 & 54.4 & 0.22 & - & - & - & - &- & 0.49\\
\midrule
MASSIVE &     [$10^9 \,M_\odot$] & [Mpc] & [$10^{-14}$]  &6 nHz& 10 nHz & 100 nHz &6 nHz& 10 nHz & 100 nHz \\%&6nHz& 10 nHz & 100 nHz \\
\midrule
NGC 7681 &  5.6 &  96.8 & 1.3 & 0.20 & 0.25 & 0.12 & 0.55 & 0.37 & 0.10\\
NGC 2693 &   4.8 & 74.4 &  1.3& 0.56 & 0.38& 0.42 & - & - & 0.18 \\%&-&-&-\\
NGC 7436 &  5.1 &  106.6 & 0.97 & 0.32& 0.43 & 0.17 & - & - & 0.15 \\%&-&-&-\\
NGC 1453 & 3.4   & 56.4 & 0.94& -& - & -& - & -& 0.20\\%&-&-&-\\
NGC 7619 & 3.2 & 54 & 0.88  & 0.34 &-& 0.30 & - & -& 0.15\\
NGC 6482 & 3.1 & 61.4 & 0.73 & 0.23 &0.31& 0.16& 0.43& 0.31& 0.09\\
NGC 0057 &  3.3 &  76.3 & 0.67 & - &-& - & - & -& 0.38\\
NGC 2832 & 3.6 & 105 & 0.57 & - &-& - & - & -& 0.80\\
NGC 5353 & 1.7 & 41.1 & 0.41 & - &-& - & - & -& 0.40\\
NGC 4555 & 2.9 & 104 & 0.39 & - &-& - & - & -& 0.35\\
NGC 6575 & 2.3 & 106 & 0.26 & - &-& - & - & -& 0.83\\
\bottomrule
\end{tabular}
%\footnotetext{measured with stars}
\end{center}
\end{table*}

%\end{savenotes}
%With the existing sky-averaged constraints on $h_0$, we can already make
% rough statements about the existence of SMBH binaries in certain galaxies which
% are sufficiently nearby and with sufficiently massive SMBHs.
As shown in Figure~\ref{fig:maxh0}, the $h_{0,\rm eq}$ values
for a subset of potential SMBH binaries are above the current \kls{all-sky} PTA upper
limits (vertical lines).  In this section we examine these {potential} sources further\kls{, as well as sources that are constrained because they are located in regions of the sky with greater PTA sensitivity, as illustrated in
%Figures 6 and 7 of 
\citet[Figure 11]{2014MNRAS.444.3709Z} and %for PPTA, and Figures 7 and 8 of
\citet[Figures 7 \& 8]{2015arXiv150902165B}.}
As discussed in Section  \ref{sec:model}, we can use the current PTA limits on continuous
GWs to obtain an upper limit on the binary mass ratio $m_2/m_1$ for each
source {using} Eq.~(\ref{ratio}). \kls{For the rest of our analysis, we focus on the limits from EPTA and PPTA, which have the strongest constraints to date. In particular, we use their constraint maps, which include the full angular dependence of these constraints. We note here that by simply using the $h_0$ constraint maps reported by these two PTAs, we are not able to constrain the binary mass ratio with a high degree of certainty. To do that properly, one would need many signal injections in the PTA data which sample the possible mass ratios and other properties of the SMBH binary. This more rigorous analysis is beyond the scope of this exploratory study. Nevertheless, we report constraints using the methods outlined above, noting that our results serve as a benchmark and as a proof of concept.}

%We note that there is a sky-dependent variation of a factor of a few in the
%PTA upper limits on the local GW strain amplitude $h_0$, as illustrated in
%%Figures 6 and 7 of 
%\citet[Figures 6 \& 7]{2014ApJ...794..141A}, %for NANOGrav, Figure 11 of
%\citet[Figure 11]{2014MNRAS.444.3709Z}, and %for PPTA, and Figures 7 and 8 of
%\citet[Figures 7 \& 8]{2015arXiv150902165B}.% for EPTA.  
%Thus more accurate constraints on
%SMBH binaries taking into account spatial variation of the GW signal will
%require a full analysis of actual PTA data, which is outside the scope
%of this paper.  Our main goal here is to perform an exploratory study of
%constraints on the mass ratios of potential SMBH binaries based on the
%sky-averaged constraints on $h_0$ reported by the three PTA teams.

We further note that despite the simple frequency scaling in Eqs.~(1) and
(4), the constraints on $h_0$ from the full analyses of
PTA data have more complicated features in the frequency dependence due to various
degeneracies between timing models and GW signals.
% that cannot be captured by the simple $f^{2/3}$ scaling
The frequency dependence of the PTA constraints on continuous GWs is
illustrated by %Figure~6 of 
\citet[Figure 6]{2014ApJ...794..141A}, %for NANOGrav,
%Figure~9 of 
\citet[Figure 9]{2014MNRAS.444.3709Z}, and % for PPTA, and Figure~6 of
\citet[Figure 6]{2015arXiv150902165B}.  All three figures show that the
constraints on $h_0$ are especially weak at frequencies corresponding to 1
and 2 inverse years, since pulsar timing systematics coming from the
earth's orbit are partly degenerate with a GW signal.  There is also an
upward trend at lower frequencies ($f \la 10$ nHz) due to the quadratic
spin-down model fit.   As shown in \citet[Figure 10]{2014MNRAS.444.3709Z},
the spectral shape of {PPTA} constraints on $h_0$ are almost exactly the same as a
function of \kls{line-of-sight} direction modulo an overall vertical shift depending on the {PTA} sensitivity \kls{as a function of angular position}. \kls{Meanwhile, the frequency dependence of EPTA constraints have a weak directional dependence.}%We can therefore convert the PTA 
%constraints on $h_0$ as a function of frequency into constraints on the
%binary mass ratios for each galaxy.  

Figure~\ref{fig:ratio_constraint} shows our resulting upper limits on the
SMBH binary mass ratio as a function of GW frequency for \kls{NGC 1600 (orange curves),} NGC 4889 (red
curves), NGC 4486 (M87; purple curves), and NGC 4649 (M60; blue curves).
Depending on the exact frequency, we obtain a tighter constraint on the mass
ratio from either EPTA (dashed curves) or PPTA (solid curves). \kls{We emphasize that we are assuming that the SMBH masses are given by their mean measured value. We anticipate that marginalizing over uncertainties on the SMBH mass (based on observational measurements) using a full analysis of the PTA data could change the constraints. This is particularly true for NGC 4889, whose mass measurement has the largest observational uncertainties. We again note that the full analysis of PTA data is beyond the scope of this exploratory study.}
%These are the three best-constrained galaxies at this
%level of approximation (i.e. by assuming sky-averaged constraints on
%$h_0$), although conveniently, these three galaxies also happen to be at
%sky locations at which the local value of $h_0$ is almost exactly the same as
%the sky-averaged $h_0$.
%as a function of the binary separation.  
%We use the sky-averaged 95\% upper limits from
%We expect that by folding in errors on SMBH mass measurements %as well as error
%estimates for local deviations from the sky-averaged value of $h_0$, the
%constraint on the mass ratio can vary by a factor of $\sim$2--3 (upwards or
%downwards).

% the constraints will deviate from the ideal scalings.  For instance,
% for both PPTA and NANOgrav the constraints on $h_0$ at 100 nHz are
% tighter than would be expected by projecting the constraints at 10 nHz
% using the simple .

In Table \ref{tab} we tabulate the upper limits on the binary black hole mass
ratios at three \kls{separate GW} frequencies (6 nHz, 10 nHz and 100
nHz) for the sources in our two samples of galaxies that can be constrained
by the current {PPTA and EPTA} limits. \kls{We chose these as representative GW frequencies which have relatively little degeneracy between a GW signal and the pulsar timing model.} We also include the total dynamical mass of the
black hole and the distance to each galaxy.
% \footnote{We also list, for completeness, the mass of M87 as measured by
%  \citet{2013ApJ...770...86W}, which is in tension with the previous
%  measurements using stellar dynamics.}.
% we report the mass ratio constraints for 16 potential SMBH binaries
%to illustrate the weak
%frequency-dependence of our constraints.  %\ref{tab:ratio}.
\footnotetext{We include the lower $\mbh$ for M87
    from \citet{2013ApJ...770...86W} based on gas rather than
    stellar kinematics. We adjust this measurement to be
    consistent with a measured distance of 16.7 Mpc. }
The top galaxies have dynamically measured black hole masses; several
additional MASSIVE galaxies do not have published dynamical measurements
yet and the total black hole mass is inferred from the $\mbh$-$\sigma$
relation.  As shown by Table \ref{tab} and Figure~\ref{fig:ratio_constraint}, we
are able to constrain the mass ratio to be less than {a few percent} for
potential binary black holes on circular orbits (within the PTA frequency
window) in several galaxies \kls{based on the constraint on $h_0$ at their positions in the sky. In spite of the fact that our results are obtained under simplifying assumptions already mentioned, we have demonstrated as a proof of concept that one can achieve percent-level constraints on SMBH binaries using existing PTA data.}

\begin{figure}%[htb]
\includegraphics[width = 0.48\textwidth]{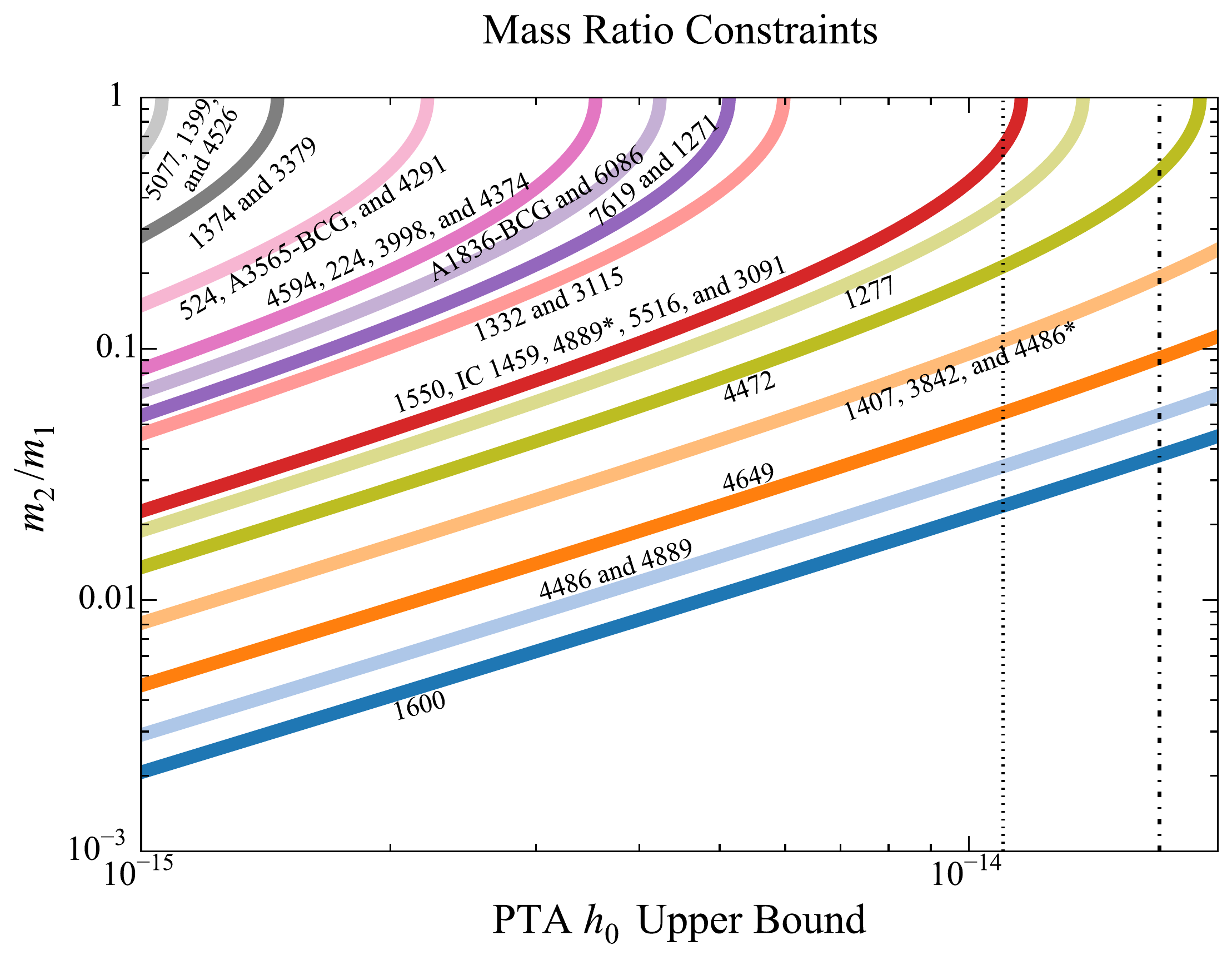}
\caption{Upper limits on the black hole mass ratio ($m_2 \le m_1$ by
  definition) of hypothetical binaries located in various galaxies as a
  function of the PTA limit on $h_0$ \kls{at the angular position of each galaxy, assuming a GW frequency of 10
  nHz}. The two vertical lines indicate the \kls{all-sky} 95\% upper limits
  on $h_0$ from EPTA (dotted; \citealt{2015arXiv150902165B}) and PPTA
  (dot-dashed; \citealt{2014MNRAS.444.3709Z}) at the same frequency.  As
  the limits on continuous GWs from PTAs become more stringent, the allowed
  binary mass ratios become smaller, and the constraints are applicable to
  more galaxies. For two galaxies, we have included two curves for each: NGC 4486
  (upper curve {denoted by an asterisk} for $\mbh$ from \citealt{2013ApJ...770...86W}; lower curve
  for $\mbh$ from \citealt{2011ApJ...729..119G}), and NGC 4889 (upper curve {denoted by an asterisk}
  for the very conservative 1$\sigma$ lower bound on $\mbh$ from
  \citealt{2011Natur.480..215M}; lower curve for their median $\mbh$).}
\label{fig:future}
\end{figure}
Upcoming continuous GW analyses of the NANOGrav 9-year data
\citep{2015arXiv150507540A} and the PPTA DR2 data
(\citealt{Shannonetal2015}) will likely place stronger limits on $h_0$ than
those compiled in this paper.
%Since the PTA limits are becoming more stringent
%and the local GW constraints vary by a factor of a few (depending on sky locations), 
In anticipation of these tighter limits, we illustrate in
Figure~\ref{fig:future} the dependence of the maximum binary mass ratio as
a function of the upper limit on $h_0$ for {galaxies with measured SMBH masses}.  This
figure illustrates future science that can be done to constrain nearby,
massive SMBH binaries by combining upcoming PTA data with dynamical measurements
of specific black holes. 
Based on Figure~\ref{fig:future}, we project that
constraints on the black holes with known masses in $\sim30$ galaxies can be obtained when the PTA
sensitivity to continuous GWs from individual sources is lowered to $h_0
\approx 10^{-15}$ \kls{at the directions in the sky where these galaxies are located}.

%Table~1 lists their
%distances and dynamically measured black hole masses.  {\bf [Elaborate.]}
Meanwhile, the galaxies that are below the current and upcoming detection
thresholds are potential GW ``hot spots'' for ongoing and future PTA
searches for continuous GWs (a topic which has been previously explored in,
for instance, \citealt{2014ApJ...784...60S} and
\citealt{2011ApJ...730...17B}). The analysis of these particular hotspots
and their implications for upcoming PTAs will be the subject of future
work.

\section{Discussion}
\label{sec:con}
%\lipsum
We have demonstrated, as a proof of concept, the efficacy of using PTA
limits on GW strain amplitudes to place constraints on individual
prospective SMBH binaries in our local volume by leveraging information
about their total dynamical masses and distances.  In particular, we have
examined two samples of galaxies and
placed upper limits on the mass ratio of SMBH binaries with sub-parsec
orbital separations in a number of galaxies. To achieve this, we have used the latest and most stringent
upper limits on $h_0$ (\kls{as a function of direction in the sky}) from PPTA and EPTA.

For several SMBHs with $\mbh \ga 5\times 10^9 M_\odot$, e.g., NGC 4889, NGC
4486, NGC 4649\kls{, and NGC 1600}, we find that the mass ratio of a hypothetical binary
would have to be less than around {a few percent}, or even as low as {a few tenths of a percent} in
some PTA frequency bands (see Figs.~\ref{fig:ratio_constraint} and
\ref{fig:future} and Table \ref{tab}).  
%we have tightly constrained the binary mass ratios.
%Although the
%results plotted in Figure \ref{fig:ratio_constraint} are a rough estimate
%given current data, they already suggest that the detectability of a SMBH binary
%in one of these three galaxies is all but ruled out. 
It is interesting to note that these limits on SMBH binary mass ratios in
massive galaxies are similar to the limit for the SMBH in the
Milky Way obtained from individual stellar motions.  The exquisite
dynamical measurements of the Galactic Center constrain a second black hole
to be less than $\sim 10^5 M_\odot$, i.e., a binary mass ratio of $\la
2.5$\% \citep{2010RvMP...82.3121G}.  \kls{PTAs are} currently not
competitive as a constraint on the SMBH at the Galactic Center due to its
low $\mbh$ (see Fig.~1) and the steep mass dependence in Eqs.~ (1) and (4).  
As we have shown, however, {PTAs are} a powerful tool for providing percent-level limits on 
the most massive SMBH binary ratios out to $\sim 100$ Mpc.

%As discussed in \citet{2013CQGra..30v4014S}, the orbits of SMBH binaries with very
%unequal masses tend to grow more elliptical over time, even if the
%eccentricity is initially low. Adding eccentricity to the orbit greatly
%suppresses the power at low frequencies, in which the GW-induced timing
%residuals are expected to be the largest. 

Although the emission of GWs tends to circularize binary orbits,
interactions of a SMBH binary with the surrounding stars and gas before it
enters the GW-driven phase can make the orbits more elliptical over time,
especially for binaries with unequal masses
\citep{1963PhRv..131..435P,2013CQGra..30v4014S,2010ApJ...719..851S,2006ApJ...651..392S,2004ApJ...611..623S,2009MNRAS.394.2255S,2011MNRAS.415.3033R,2007ApJ...671...53M,2007ApJ...656..879M,1996NewA....1...35Q,2011ApJ...732L..26P,2005ApJ...634..921A,2009MNRAS.393.1423C}.
At the characteristic eccentricities quoted by these studies, the detection
power of PTAs at low frequencies (where GW-induced timing residuals are
expected to be largest) is greatly suppressed
\citep{2007PThPh.117..241E,2004ApJ...611..623S,2011MNRAS.411.1467K}.
Therefore, even if there is a very unequal-mass SMBH binary in \kls{NGC 1600,} NGC 4889,
NGC 4486, or NGC 4649, its orbit will likely be highly eccentric and
render a detection improbable in the near future. \kls{Upcoming analyses based on the formalism developed in \citet{2016ApJ...817...70T} could shed light on this issue further. }%Analyses of PTA
%data taking into account local variations in the limit on $h_0$ could
%sharpen this statement.
% if this kind of analysis is performed on future PTA data,
%the detectability of SMBH binaries in the local universe could soon be quite
%tightly constrained.

%While local variations in the constraints on $h_0$ can change our mass
%ratio results, even the worst-case-scenario of having a SMBH binary in the lowest
%region of PTA sensitivity would only weaken our limits by a factor of
%$\sim 2$--3. Conversely, the local constraints on $h_0$ can be stronger than
%the sky-averaged value used in this paper by a factor of a few.  For the
%galaxies located in these areas of the sky in our study, we expect the
%limits obtained in the paper to be tightened.  This is particularly true
%for SMBHs in the MASSIVE galaxies, several of which are along similar lines
%of sight as the PTA pulsars.

{Based on the benchmarks that we have set here,} we anticipate that a full analysis of existing or upcoming PTA data
will {prove extremely useful by providing more accurate} constraints on the SMBH binaries in our
sample galaxies. \kls{By searching the full parameter space of binary mass ratios, binary inclination angles, etc., such analysis would provide meaningful limits on the astrophysics of SMBH binaries and their relationship with the host galaxies. We further anticipate that such analysis would be useful to incorporate into searches for anisotropy in the gravitational wave background (such as the one performed in \citealt{2015PhRvL.115d1101T}).} New dynamical measurements of $\mbh$ from the MASSIVE
Survey will also remove the uncertainties associated with using the $\mbh$-$\sigma$
relation as a mass proxy and provide more realistic constraints for
potential binaries in those galaxies. 

%\kls{It may be unsurprising that we have constrained the existence of nearby binaries through the non-detection of continuous gravitational waves, given the fact that the GW background is much more likely to be detected first \citep{2015MNRAS.451.2417R}. However, the non-detection of the GW background \citep{%2013Sci...342..334S,
%  2015arXiv150803024A, 2015MNRAS.453.2576L,Shannonetal2015} is already beginning to challenge the paradigm for how we think about SMBHB assembly and interactions between the binary and the environment.  Further limits from constraining extremely massive, nearby binaries could bolster the upheaval of the usual SMBHB models. The exclusion of SMBHBs in our local volume would have sweeping implications for our understanding of SMBHs in a cosmological context and would provide new avenues of inquiry about their roles in affecting the properties of host galaxies.}

%The exclusion of SMBH binaries in our local volume would have sweeping
%implications for our understanding of SMBHs in a cosmological
%context. Already, bounds on the stochastic GW background from
%\citet{Shannonetal2015} are challenging the paradigm for how we think about
%SMBH binary assembly and limits on SMBH binaries from continuous GWs could bolster this
%upheaval.

\section*{Acknowledgements}
It is an immense pleasure to thank Justin Ellis, Matthew Kerr, Adrian Liu, Nicholas McConnell, Scott Ransom, Leo Stein, and Xingjiang Zhu for useful conversations and correspondence pertaining to this work. \kls{We also wish to thank Stanislav Babak and Xingjiang Zhu (on behalf of EPTA and PPTA, respectively) for providing us with their constraint maps. Finally, we thank our referee, Alberto Sesana, for his detailed comments on the original version of the manuscipt.} KS is
supported by a Hertz Foundation Fellowship and by a National Science
Foundation Graduate Research Fellowship. CPM is supported in part by National Science Foundation
grant AST-1411945.

\bibliographystyle{mnras}
\bibliography{pta_v9}

\label{lastpage}

\end{document}